\documentclass[onecolumn,aps,prb,superscriptaddress,floatfix,10pt,altaffilletter,showpacs]{revtex4-1}
\usepackage{amsmath}
\usepackage{amssymb,amsfonts}
\usepackage{mathrsfs}
\usepackage{bbm}
\usepackage{bm}
\usepackage{dcolumn}
\usepackage{dsfont}
\usepackage[english]{babel}
\usepackage{ae}
\usepackage{url}
\usepackage[dvips]{graphicx}
\usepackage{hyperref}
\newcommand{\braket}[3][]{\left\langle{#2}\middle\vert{#1}\middle\vert{#3}\right\rangle}
\newcommand{\waa}[1]{\omega_{\alpha_{#1}\alpha_{#1}^\prime}}
\newcommand{\waan}{\omega_{\alpha_{1}\alpha_{2}}}
\newcommand{\waap}{\omega_{\alpha_{1}^\prime\alpha_{2}^\prime}}
\newcommand{\LL}{\mathcal{L}}
\newcommand{\HH}{\mathcal{H}}
\newcommand{\AAA}{\mathcal{A}}
\newcommand{\KK}{\mathcal{K}}
\newcommand{\PP}{\mathcal{P}}

\newcommand{\Pp}{\mathbb{P}}
\newcommand{\Ll}{\mathbb{L}}
\newcommand{\RR}[1]{\Re \left( #1 \right)}

\newcommand{\aaaa}{\alpha_1\alpha_2,\alpha_1^\prime\alpha_2^\prime}
\newcommand{\tr}[2][]{\mathrm{tr}_{#1}\left\{#2\right\}}
\newcommand{\ener}{\epsilon}

\renewcommand{\tilde}{\widetilde}
\newcommand{\ud}{\mathrm{d}}
\newcommand{\rhov}{\varrho}

\begin{document}

\author{M. Pepe}
\affiliation{Dipartimento di Fisica, Politecnico di Torino,
Corso Duca degli Abruzzi 24, 10129 Torino, Italy}

\author{D. Taj}
\affiliation{Centre de Physique Théorique, Campus de Luminy, 13288 Marseille, France}

\author{R.C. Iotti}
\affiliation{Dipartimento di Fisica, Politecnico di Torino,
Corso Duca degli Abruzzi 24, 10129 Torino, Italy}

\author{F. Rossi}
\affiliation{Dipartimento di Fisica, Politecnico di Torino,
Corso Duca degli Abruzzi 24, 10129 Torino, Italy}

\title{Microscopic theory of energy dissipation and decoherence in solid-state systems: A reformulation of the conventional Markov limit}

\begin{abstract}
We present and discuss a general density-matrix description of energy-dissipation and decoherence phenomena in open quantum systems, able to overcome the intrinsic limitations of the conventional Markov approximation. In particular, the proposed alternative adiabatic scheme does not threaten positivity at any time. The key idea of our approach rests in the temporal symmetrization and coarse graining of the scattering term in the Liouville-von Neumann equation, before applying the reduction procedure over the environment degrees of freedom. The resulting dynamics is genuinely Lindblad-like and recovers the Fermi's golden rule features in the semiclassical limit.
Applications to the prototypical case of a semiconductor quantum dot exposed to incoherent phonon excitation peaked around a central mode are discussed, highlighting the success of our formalism with respect to the critical issues of the conventional Markov limit.
\end{abstract}

\maketitle

\section*{Introduction}

In spite of the quantum-mechanical nature of the electron and photon dynamics in the core region of typical
solid-state nanodevices, the overall behavior of such quantum systems is often characterized by a non-trivial
interplay between phase coherence and energy relaxation/dephasing phenomena.\cite{RMP,FaustoBook}
This fundamental topic is getting more and more critical in the last years, due to the continuous downscaling
in size that state-of-the-art technology allows for. The proper modelling of, e.g., charge-carrier dynamics
in cascade nanodevices,\cite{capasso,ROPP} rather than of spin/charge-based implementations of
quantum-computation algorithms in solid-state logic gates,\cite{QLG1,QLG2} are just a couple of
examples demanding for a unitary microscopic description. Not to forget the still unsolved issue represented
by the effects due to the presence of spatial boundaries (contacts).\cite{Frensley}

Contrarily to purely atomic and/or photonic systems --where decoherence phenomena may be successfully
described via phenomenological adiabatic-decoupling procedures-- the quantitative modelling of realistic
solid state devices requires to account for both coherent and incoherent processes, on equal footing.
To this aim, motivated by the power and flexibility of the semiclassical kinetic theory \cite{JL}
in describing a large variety of interaction mechanisms, a quantum generalization of the standard
Boltzmann collision operator has been proposed;\cite{RMP} the latter, obtained via the conventional
Markov (CM) limit, describes the evolution of the reduced density matrix in terms of in- and
out-scattering super-operators. However, contrary to the semiclassical case, such collision super-operator
does not preserve the positive-definite character of the density-matrix operator because of its
non-temporal symmetric construction.

This serious limitation was originally noted by Spohn and co-workers~\cite{D-S} three decades ago;
in particular, they clearly pointed out that the choice of the adiabatic decoupling strategy is definitely
not unique, and only one among the diverse possibilities, developed by Davies in
Ref.s~[\onlinecite{QOS}],[\onlinecite{Davies}], could be shown to preserve positivity: it was the case of a ``small'' subsystem
of interest interacting with a thermal environment, and selected through a partial
trace reduction. Unfortunately, the theory was restricted to finite-dimensional subsystems only
(i.e., $N$-level atoms), and to the particular projection scheme of the partial trace.

Inspired by the pioneering papers by Davies and co-workers, aim of the present work is to propose an
alternative and more general adiabatic procedure. In particular, we shall further explore the path
recently proposed in Ref.s~[\onlinecite{tajpra}],[\onlinecite{tajepj}] which, in the discrete-spectrum case reduces to Davies' model,\cite{Davies} for diagonal states gives the well known Fermi's golden rule,\cite{Alicki,FGR} always has an explicit temporal symmetry, and, finally, describes a genuine Lindblad evolution~\cite{Lindblad} even in the infinite-dimensional/continuous spectrum case. Application of the latter to a prototypical case of interest will be discussed, showing up, in contrast to the above mentioned pathologies of the CM approach, a reliable treatment of energy-dissipation and dephasing processes in semiconductor
quantum devices.

Contrary to standard master-equation formulations,\cite{Davies,MED} our approach leads to equations
that are always of Lindblad type,\cite{Lindblad} so that positivity is native in our
adiabatic-decoupling strategy.
Moreover, our approximation scheme holds true under the same validity regime
of the conventional Markov approach: the so-called weak-coupling limit,\cite{Davies2} where the subsystem density matrix in the interaction frame moves slowly with respect to perturbative effects.

The present article is organized as follows. After a general introduction, in Section~\ref{GF} the key features of the formal CM approach and of the proposed alternative treatment are recalled and compared. Section~\ref{App} presents the application of both schemes to a prototypical system, with the purpose of analyzing the failure of the CM and the success of the novel approach. The main conclusions, together with a brief summary, are contained in Section~\ref{s-SC}.

\section{General framework}
\label{GF}
To be more precise about the main features of the problem, it could be useful to recall its general
formulation based on the fully operatorial approach proposed in Ref.~[\onlinecite{PRB}].
Given a generic physical observable $A$ ---described by the operator ${\hat A}$--- its quantum plus
statistical average value is given by $A = {\rm tr}\left\{{\hat A} {\hat \rho}\right\}$,
where ${\hat \rho}$ is the so-called density-matrix operator. Its time evolution is dictated
by the global (system plus environment) Hamiltonian, that can be regarded as the sum of a
noninteracting contribution, plus a system-environment coupling
term: ${\hat H} = {\hat H}_\circ + {\hat H}'$; the corresponding equation of motion for
the density-matrix operator ---also known as Liouville-von Neumann equation--- in
the interaction picture is given by:
\begin{equation}\label{LvN_i}
\frac{\mathrm{d}{\hat \rho}^i}{\mathrm{d} t}(t) = -i \left[\hat{\cal H}^i(t),
{\hat \rho}^i\right] \, ,
\end{equation}
where $\hat{\cal H}^i$ denotes the interaction Hamiltonian $\hat{H}'$ written in units of $\hbar$.

The key idea, common to any perturbation approach, is that the effect of ${\hat H}'$ is ``small'' compared to the free evolution dictated by ${\hat H}_\circ$. Following this spirit, by formally
integrating Eq.~(\ref{LvN_i}) from $t_\circ$ to the current time $t$, and inserting
such formal solution for ${\hat \rho}^i(t)$ on the right-hand side of Eq.~(\ref{LvN_i}),
we obtain an integro-differential equation of the form:
\begin{equation}\label{IDE}
\frac{\mathrm{d} {\hat \rho}^i}{\mathrm{d} t}(t)\! =\! -i \left[\hat{\cal H}^i(t),
{\hat \rho}^i(t_\circ)\right]\! -\! \int_{t_\circ}^t\! dt'\!
\left[\hat{\cal H}^i(t), \!\left[\hat{\cal H}^i(t'), {\hat
\rho}^i(t')\right]\right].
\end{equation}
So far, no approximation has been introduced: Eq.s~(\ref{LvN_i}) and~(\ref{IDE})
are fully equivalent, we have just isolated the first-order contribution from the
exact time evolution in Eq.~(\ref{LvN_i}).

The final goal within the Markovian picture is an equation in which the time evolution of the complete density matrix, in Schr\"odinger representation and omitting the coherent rotation due to the first order contribution, can be expressed by means of a super-operator $\mathbb{L}^{global}$ as
\begin{equation}
\label{SchGen}
\frac{\ud \hat{\rho}}{\ud t}= \mathbb{L}^{global}(\hat{\rho})\, .
\end{equation}
In particular, by denoting with $\{\vert \lambda \rangle\}$ the eigenstates of the noninteracting
Hamiltonian $\hat{H}_\circ$, and neglecting energy-renormalization
contributions,\cite{PRB} the effective equation~(\ref{SchGen}) written in such a
basis has the form:
\begin{equation}\label{LvN-eff-lambda}
\frac{\mathrm{d}\rho_{\lambda_1\lambda_2}}{\mathrm{d} t} =
\frac{1}{2} \sum_{\lambda'_1\lambda'_2}
\left[{\cal P}_{\lambda_1\lambda_2,\lambda'_1\lambda'_2}
\rho_{\lambda'_1\lambda'_2} - {\cal P}_{\lambda_1\lambda'_2,\lambda'_1\lambda'_1}
\rho_{\lambda'_2\lambda_2} \right] + {\rm h.c.}
\end{equation}

Equation~(\ref{LvN-eff-lambda}) describes the evolution of
the global --system ($S$) plus environment ($E$)-- density matrix.
However, in the study of electronic quantum phenomena in semiconductor nanostructures most of the physical quantities of interest depend on the electronic system only; the corresponding operators have therefore the form
$\hat{O} = \hat{O}^S \otimes \hat{\mathbbm{1}}^E$.
Their expectation values may then be written as
\begin{eqnarray} \label{average3}
\langle \hat{O} \rangle = \tr{\hat{O}\hat{\rho}}=
\tr{(\hat{O}^S \otimes \hat{ \mathbbm{1}}^E)\hat{\rho}}=
\tr{\hat{O}^S \hat{\rho}^S}
\end{eqnarray}
where
\begin{eqnarray} \label{parziale}
\hat{\rho}^S= \tr[E]{\hat{\rho}},
\end{eqnarray}
is the reduced or electronic density matrix, obtained by tracing over the environment variables.

The above mentioned reduction scheme can be interpreted in terms of a projection super-operator $\Pp$, that acts on the generic global density matrix performing the partial trace over the
environment degrees of freedom:
\begin{equation}
\label{fact}
\Pp {\hat{\rho}}=
\tr[E]{\hat{\rho}} \otimes  \hat{\rho}^E= \hat{\rho}^S\otimes \hat{\rho}^E \, ,
\end{equation}
$\hat{\rho}^{E}$ being the environment density-matrix operator.
This step is crucial to study the weak-coupling limit; as we shall see, it needs to be considered explicitly only later on, when talking about positivity. For the moment, we limit ourselves to assume that $\Pp\left([\hat H'(t),\Pp {\hat \rho}]\right)= 0$: This is physically justified for a wide family of perturbing Hamiltonians,\cite{Davies} and allows us to neglect the first order term on the right hand side of Eq.~(\ref{IDE}), that gives no contribution when projected.\cite{renorm} Hence for ease of exposition we shall now study all the projection-independent features of our model, safely neglecting first order terms.

Since $\hat{\rho}^S$ is the only quantity entering the evaluation of the average value in
Eq.~(\ref{average3}), it is desirable to derive a corresponding
equation of motion for the reduced density-matrix operator.
However, $\Pp$ does not commute with the scattering super-operator one wants to project, thus hampering us in obtaining a closed equation of motion for $\hat{\rho}^S$. Once at this point, to overcome such a problem, the typical procedure is to assume that the dynamics of the electronic subsystem does not significantly perturb the environment, which therefore is described by a time-independent density matrix. This can be achieved when the latter is, for example, maintained in a quasi-equilibrium regime and/or driven by some external mechanism.

Let us now apply this approach to Eq.~(\ref{SchGen}), in the framework of Eq.~(\ref{fact}):
\begin{gather}
\label{projGeneral}
\frac{\mathrm{d} \hat{\rho}}{ \mathrm{d}t }
= \mathbb{L}^{global}(\hat{\rho})\;
\rightarrow
\Pp\frac{\mathrm{d} (\Pp{\hat{\rho}})}{ \mathrm{d}t }
= \Pp\mathbb{L}^{global}(\Pp\hat{\rho})\, .
\end{gather}

Rewriting the latter on the basis given by the unperturbed electronic states $\{\vert \alpha \rangle \}$, one obtains
\begin{eqnarray} \label{matrice}
\frac{\mathrm{d} \rho^S_{\alpha_{1} \alpha_{2}}}{\mathrm{d} t}= \sum_{\alpha_{1}'\alpha_{2}'}
\Ll_{\alpha_{1}\alpha_{2},\alpha_{1}'\alpha_{2}'}
^{\rm scatt}
\,
\rho^S_{\alpha_{1}'\alpha_{2}'}
\end{eqnarray}
where the complete super-operator
\begin{equation}
\Ll
^{\rm scatt}
=\tr[E]{\Ll^{global}(\hat{\rho}^S \otimes \hat{\rho}^E)}
\end{equation}
is made up of four terms, that is
\begin{equation}
\label{defLL}
\Ll_{\alpha_{1}\alpha_{2},\alpha_{1}'\alpha_{2}'}
^{\rm scatt}
=\frac{1}{2}\!\left(\!
\PP_{\alpha_{1}\alpha_{2},\alpha_{1}'\alpha_{2}'}\! - \!\!
\sum_{\alpha_{3}}
\PP_{\alpha_{1}\alpha_{1}',\alpha_{3}\alpha_{3}}\,\delta_{\alpha_{2}'\alpha_{2}}
\! + {\rm h.c.} \! \right) \, .
\end{equation}
Here, the generalized scattering rates $\PP$ are obtained applying the projection $\Pp$ to
Eq.~(\ref{LvN-eff-lambda}). In this respect, diverse approaches may be employed.
In particular, in the following Subsection~\ref{NewM} the generalized scattering rates $\PP$ will derived according to an alternative procedure for the Markov limit, able to overcome the well-known critical issues of the conventional one. To better compare the key features of both schemes, Subsection~\ref{CM} contains a brief description of the latter.

\subsection{Conventional Markov limit}
\label{CM}%

Let us therefore focus on the time integral in Eq.~(\ref{IDE}). Here, the two quantities to be integrated over $t'$ are the interaction Hamiltonian $\hat{\cal H}^i$ and the density-matrix operator ${\hat \rho}^i$. In the spirit of the perturbation
approach previously recalled, the time variation of ${\hat \rho}^i$ can be considered
adiabatically slow compared to that of the Hamiltonian $\hat{\cal H}$ written in the
interaction picture, i.e.,
$\hat{\cal H}^i(t') = {\hat U}^\dagger_\circ(t') \hat{\cal H} {\hat U}^{ }_\circ(t')$;
indeed, the latter exhibits rapid oscillations due to the noninteracting evolution operator
${\hat U}_\circ(t) = e^{-\frac{i{\hat H}_\circ t}{\hbar}}$.
Therefore, in the standard (and problematic) Markov approximation the
density-matrix operator ${\hat \rho}^i$ in the interaction frame is simply taken out of the time integral and evaluated at the current time $t$.

Following such a prescription, the system dynamics written
in the Schr\"odinger picture for the case of a time-independent interaction
Hamiltonian $\hat{\cal H}$ comes out to be:
\begin{equation}\label{LvN-eff}
\frac{\mathrm{d}\hat{\rho}}{\mathrm{d} t} = -i[\hat{\HH} , \hat{\rho}]
-\frac{1}{2} \left[\hat{\cal H},
\left[\hat{\cal K},{\hat \rho}\right]\right]
\equiv
-i[\hat{\HH} , \hat{\rho}] +
\mathbb{L}^{global}_{CM}(\hat{\rho})
\end{equation}
with
\begin{equation}\label{calK}
\hat{\cal K}
=
2 \int_{t-t_\circ}^{0} \mathrm{d}t'
\hat{\cal H}^i(t')
=
2 \int_{t-t_\circ}^{0} \mathrm{d}t' {\hat U}^{\dagger}_\circ(t')
\hat{\cal H} {\hat U}_\circ(t') \ .
\end{equation}
The effective equation in~(\ref{LvN-eff}) has still the double-commutator structure in~(\ref{IDE})
but it is now local in time. In this scheme, the generalized scattering rates of the
Eq.~(\ref{LvN-eff-lambda}), obtained within the completed-collision limit $t_\circ \to -\infty$, turn out to be
\begin{equation}
\label{calP}
{\cal P}_{\lambda_1\lambda_2,\lambda'_1\lambda'_2} = \frac{2\pi}{\hbar} H'_{\lambda_1\lambda'_1} H^{\prime *}_{\lambda_2\lambda'_2} \delta(\epsilon_{\lambda_2} - \epsilon_{\lambda'_2}) \ ,
\end{equation}
where $\epsilon_\lambda$ denotes the energy corresponding to state $\vert \lambda \rangle$.

Once projected with $\Pp$, and considering a density matrix in the form of Eq.~(\ref{fact}), Eq.~(\ref{LvN-eff}) reduces to
\begin{eqnarray}
\label{projCM}
\frac{\mathrm{d}}{\mathrm{d} t} \hat{\rho}^S=-i[\hat{\HH}_\circ , \hat{\rho}^S]
- \frac{1}{2}\Pp([\hat{\HH}',[\hat{\KK},\hat{\rho}^S]]) \, .
\end{eqnarray}

The Markov limit recalled so far leads to significant modifications in the system dynamics:
while the exact quantum-mechanical evolution in Eq.~(\ref{LvN_i}) corresponds to a fully
reversible and isoentropic unitary transformation, the instantaneous double-commutator
structure in Eq.~(\ref{LvN-eff}) describes, in general, a non-reversible (i.e., non unitary)
dynamics characterized by energy dissipation and dephasing. However, since any effective
Liouville super-operator should correctly describe the time evolution of $\hat\rho$ and
since the latter, by definition, needs to be trace-invariant and positive-definite at any
time, it is imperative to determine if the Markov super-operator in Eq.~(\ref{LvN-eff})
fulfills this two basic requirements.

As far as the first issue is concerned, in view of its commutator structure, it is easy
to show that this effective super-operator is indeed trace-preserving. In contrast,
as discussed extensively in Ref.~[\onlinecite{PRB}], the latter does not ensure that
for any initial condition the density-matrix operator will be positive-definite at any time.
This is by far the most severe limitation of the CM approximation.

The well-known semiclassical or Boltzmann theory~\cite{JL} can be easily derived
from the quantum-transport formulation presented so far, by introducing the
so-called diagonal or semiclassical approximation. The latter consists in
neglecting all non-diagonal density-matrix elements (and therefore any quantum-mechanical
phase coherence between the generic states $\lambda_1$ and $\lambda_2$), i.e.,
$\rho_{\lambda_1\lambda_2} = f_{\lambda_1} \delta_{\lambda_1\lambda_2}$,
where the diagonal elements $f_\lambda$ correspond to the semiclassical distribution
function over our noninteracting basis states. Within such approximation scheme,
the quantum-transport equation~(\ref{LvN-eff}) reduces to the
well-known Boltzmann equation:
\begin{equation}\label{BTE}
\frac{\mathrm{d} f_\lambda }{\mathrm{d} t} =
\sum_{\lambda'} \left(
P_{\lambda\lambda'} f_{\lambda'} - P_{\lambda'\lambda} f_\lambda
\right)\ ,
\end{equation}
where
\begin{equation}\label{P}
P_{\lambda\lambda'} = {\cal P}_{\lambda\lambda,\lambda'\lambda'} =
\frac{2\pi}{\hbar} |H'_{\lambda\lambda'}|^2 \delta\left(\epsilon_{\lambda}-\epsilon_{\lambda'}\right)
\end{equation}
are the conventional semiclassical scattering rates given by the well-known Fermi's golden rule.\cite{FGR}

At this point it is crucial to stress that, contrary to the non-diagonal
density-matrix description previously introduced, the Markov limit combined
with the semiclassical or diagonal approximation ensures that at any time $t$
our semiclassical distribution function $f_\lambda$ is always positive-definite.
This explains the ``robustness'' of the Boltzmann transport equation in~(\ref{BTE}),
and its extensive application in solid-state-device modeling as well as in many other fields, where quantum effects play a very minor role. In contrast, in order to
investigate genuine quantum-mechanical phenomena, the corresponding CM super-operator $\mathbb{L}^{global}_{CM}$ obtained from Eq.~(\ref{LvN-eff})
cannot be employed, since it does not preserve the positive-definite
character of the density matrix $\rho_{\lambda_1\lambda_2}$.

As highlighted before, the origin of these pathologies is in the lack of temporal symmetry in the construction of the CM approximation; namely, in the double commutator expression in Eq.s~(\ref{LvN-eff}) and~(\ref{projCM}). Motivated by this evidence, in the next Subsection, we shall propose an alternative adiabatic approximation method in which the hamiltonian operators of Eq.~(\ref{IDE}) are always time-balanced, so that the temporal symmetry remains always explicit and is never broken.

\subsection{Alternative approach to the Markov limit}
\label{NewM}%

To introduce our alternative formulation of the problem,
let us go back to the integro-differential equation in~(\ref{IDE}). We can formally integrate the latter from $t_\circ$ to $t$, and obtain
\begin{equation} \label{NakaZwanzig}
\hat\rho^i (t)=\hat\rho^i (t_\circ) -i \int_{t_\circ}^t \!\! \mathrm{d} t_1 \!
\left[\hat{\cal H}^i(t_1),
{\hat \rho}^i(t_\circ)\right]
- \int_{t_\circ}^t\!\! \mathrm{d} t_1 \! \int_{t_\circ}^{t_1}\!\! \mathrm{d} t_2\!
\left[\hat{\cal H}^i(t_1), \!\left[\hat{\cal H}^i(t_2), {\hat
\rho}^i(t_2)\right]\right]\, .
\end{equation}
This equation, named after Sadao~Nakajima and Robert~Zwanzig,
is still exact, and it is nothing but the Dyson
equation after two iterations.

As far as the first integral on the right-hand side of Eq.~(\ref{NakaZwanzig}) is concerned, the most it can cause is an energy renormalization effect that, after the partial-trace projection in our proposed scheme, will straightforwardly result into a coherent term. Since the latter does not affect positivity in any case, we shall forget about it in the following.

Let us now perform a change of variables. In particular, starting from the two times $t_1$ and $t_2$, we introduce a ``relative'' time
\begin{subequations}
\begin{align}
\tau = t_1-t_2 \, ,
\end{align}
and a ``macroscopic'' time
\begin{align}
T = \frac{t_1+t_2}{2} \, .
\end{align}
\end{subequations}

This change of variables has very sound bases, as it is typical and well
established for a wide variety of contests, such as Wigner's phase-space
formulation of quantum mechanics,\cite{wigner} standard quantum
kinetics Green's function methods (see e.g., [\onlinecite{haug,haug2}])
and even classical radiation theory (e.g., in the treatment of Bremsstrahlung).
The basic idea is that the relevant time characterizing/describing our effective
system evolution is the macroscopic time $T$.

Having this profound difference between $T$ and $\tau$ in mind, we can rewrite Eq.~(\ref{NakaZwanzig}) as:
\begin{equation} \label{NakaZwanzig-2}
\hat\rho^i (t) = \hat\rho^i (0)
- \int\limits_{0}^{t}\!\! \mathrm{d} T \!\!
\int\limits_{0}^{g(t,T)}\!\!\!\!\! \mathrm{d} \tau
\left[\hat{\cal H}^i\left(T+\frac{\tau}{2} \right),
\left[\hat{\cal H}^i\left(T-\frac{\tau}{2} \right),
{\hat{\rho}}^i\left(T - \frac{\tau}{2} \right)\right]\right] \, ,
\end{equation}
where the function $g$ comes from the domain
of integration in the $(T,\tau)$ plane and
can be expressed as~\cite{tajAHP}
\begin{eqnarray} \label{g}
g(t,T)=
\left|\frac{t}{2} - \left|T- \left(\frac{t}{2} \right)
\right| \right| \approx \bar{t} \, .
\end{eqnarray}

The integral over the relative time $\tau$ in Eq.~(\ref{NakaZwanzig-2}) requires some manipulation. In particular, by introducing a proper gaussian cut-off factor,
$e^{-\frac{\tau^2}{2 {\overline{t}}^2}}$,
the upper limit $g(t,T)$ may be extended to infinity; that is
\begin{eqnarray}
\label{approx1}
\int_{0}^{g(t,T)}\!\!\! f(\tau) \,\mathrm{d} \tau
\approx
\int_{0}^{\infty}\!\!
f(\tau) \,e^{-\frac{\tau^2}{2 \overline{t}^2} } \, \mathrm{d} \tau \,.
\end{eqnarray}
In other words, the last integral on $\tau$ gets effectively cut off at the collision time $\overline t$, having in mind that the contributions resulting from the double commutator structure get negligible at longer times. In particular, $\overline{t}$ scales up as some negative power of the coupling constant (that is the ratio between the interaction matrix element and the unperturbed energy gap) and goes to infinity when the interaction is weak.

In this framework, we now evaluate the first derivative of Eq.~(\ref{NakaZwanzig-2}) with respect to $t$ and obtain:
\begin{equation} \label{IDE-new}
\frac{\mathrm{d}}{\mathrm{d} t } \hat\rho^i (t)=
-\int_{0}^{\infty}\!\! \mathrm{d} \tau\, e^{-\frac{\tau^2}{2 \overline{t}^2} }
\left[\hat{\cal H}^i\left(t+\frac{\tau}{2} \right),
\left[\hat{\cal H}^i\left(t-\frac{\tau}{2} \right),
{\hat \rho}^i\left(t - \frac{\tau}{2} \right)\right]\right] \, .
\end{equation}
In the spirit of the adiabatic approximation previously recalled,
the density-matrix operator ${\hat \rho}^i$, being a slowly varying function of
$\tau$, can be taken out of the time integral and evaluated at the current time $t$.

The skew-adjoint part of Eq.~(\ref{IDE-new}) --the so-called scattering part-- is our main interest, the self-adjoint part being just an energy-renormalization term that
does not threaten positivity; we then get:
\begin{equation} \label{LvN-eff-new1}
\frac{\mathrm{d}}{\mathrm{d} t} {\hat \rho}^i(t) = -\frac{1}{2}
\int_{-\infty}^\infty \mathrm{d}\tau\; e^{-\frac{\tau^2}{2{\overline{t}}^2}}
\left[\hat{\cal H}^i\left(t+ \frac{\tau}{2}
\right), \left[\hat{\cal H}^i\left(t- \frac{\tau}{2}\right),
{\hat \rho}^i\left(t\right)\right]\right]\, .
\end{equation}
It is evident how the proposed time symmetrization gives rise to a
fully symmetric super-operator, contrarily to the strongly asymmetric
Markov super-operator in Ref.~[\onlinecite{Davies2}].

The second crucial step to get a genuine Lindblad
super-operator for the global dynamics is to exploit once again the
slowly-varying character of the density-matrix operator
$\hat\rho^i$ on the right-hand side of Eq.~(\ref{LvN-eff-new1}).
The key idea is to perform on both sides of
Eq.~(\ref{LvN-eff-new1}) a so-called temporal ``coarse graining'',
i.e., a weighted time average on a microscopic scale where the variations of $\hat\rho^i(t)$ are negligible.
In the weak and intermediate coupling regime this time-scale is comparable to $\overline{t}$; we shall then perform the above mentioned convolution employing once again a gaussian correlation function of width
$\frac{\overline{t}}{2}$, i.e.,
\begin{equation}
\label{LvN-eff-new3}
\frac{\mathrm{d}}{\mathrm{d} t}\, {\hat \rho}^i (t) \, = \,
- \int_{-\infty}^\infty \!\!\! \mathrm{d}\tau'\:
\frac{e^{-\frac{{4\tau'}^2 }{2\overline{t}^2}}} {\sqrt{2\pi}\,\bar{t}}
\!\int_{-\infty}^\infty \!\!\! d\tau\: e^{-\frac{\tau^2}{ 2\overline{t}^2}}
\! \left[\hat{\cal H}^i\left(t\!-\!\tau'\!+\! \frac{\tau }{2} \right),
\left[\hat{\cal H}^i\left(t\!-\!\tau'\!-\! \frac{\tau}{2}\right),
{\hat \rho}^i\left(t\right)\right]\right].
\end{equation}
Moving back to the original Schr\"odinger picture and combining
the two gaussian distributions, the above equation can be
rewritten\cite{Note-6} in the following compact form:
\begin{equation}\label{Lindblad-bis}
\frac{\mathrm{d} {\hat \rho}} {\mathrm{d} t} = -\frac{1}{2} \left[\hat{\cal L},
\left[\hat{\cal L}, {\hat \rho}\right]\right]
\end{equation}
with\cite{Note-7}
\begin{equation}\label{calL-new}
\hat{\cal L} = \left(
\frac{1}{\sqrt{2\pi}\,\bar{t}}\right)^\frac{1}{2}
\int_{-\infty}^\infty dt' \;\hat{\cal H}^i(t')\; e^{-\frac{{t'}^2}{ 4\overline{t}^2}} \ .
\end{equation}
This is the genuine Lindblad-like super-operator we were looking
for; indeed, the operators $\hat{\cal L}$ are always Hermitian, and therefore the
effective dynamics they induce is positive-definite.

Let us finally rewrite the new Markov super-operator in Eq.~(\ref{Lindblad-bis}) in our noninteracting basis $\{ \vert \lambda \rangle \}$,
defined by the (possibly generalized) eigenvectors of $H_\circ$. In this way, we now obtain an effective equation of motion in the same form of Eq.~(\ref{LvN-eff-lambda})
with symmetrized quantum scattering rates
\begin{equation}
\label{calPtilde}
{\cal P}_{\lambda_1\lambda_2,\lambda'_1\lambda'_2} =
\frac{2\pi}{\hbar} H'_{\lambda_1\lambda'_1} H^{\prime
*}_{\lambda_2\lambda'_2} {\frac{1}{\sqrt{2\pi} \overline\epsilon}
\exp{
\left\{
- \frac{ \left( \epsilon_{\lambda_1}-\epsilon_{\lambda'_1} \right)^2
		+ \left( \epsilon_{\lambda_2}-\epsilon_{\lambda'_2} \right)^2  }
		{4 \overline{\epsilon}^2 }
\right\} } } \, .
\end{equation}
Here, $\overline{\epsilon} = \hbar/\,\overline{t}=\hbar \overline{\omega}$ is a measure of the energy uncertainty in the interaction process induced by our temporal coarse graining.

The above scattering super-operator can be regarded as a generalization of the conventional Fermi's
golden rule to the density matrix formalism; Indeed, in the semiclassical diagonal case
($\lambda_1=\lambda_2,\lambda_1'=\lambda_2'$) it reduces
to what could be considered a dressed vertex-smoothed version of the Fermi's golden rule
\begin{equation}
P_{\lambda\lambda'} =
{\cal P}_{\lambda\lambda,\lambda'\lambda'} =
\frac{2\pi}{\hbar}\, |H'_{\lambda\lambda'}|^2\,
 \frac{1}{\sqrt{2\pi} \overline\epsilon} \,
e^{- \left(\epsilon_{\lambda}-\epsilon_{\lambda'} \right)^2/2\overline{\epsilon}^2} \, .
\end{equation}
In the limit of infinite correlation-time
($\overline{\omega} \to 0$), the
standard scattering rates given by Eq.~(\ref{P}) are readily recovered.

At this point, one may wonder whether the approximate dynamics so obtained suffers from lack of generality, the gaussian smoothing and cut-off having been put by hand. However, it is important to stress that one has to limit the analysis to the asymptotic features, as the approximation is valid in the weak-coupling regime.
As such, and without loss of generality, our gaussian choice, among all the possible asymptotic markovian approximations of the exact dynamics that guarantee a positive evolution, is a good representative. Moreover the use of gaussian functions has the marked advantage of allowing for the analytical treatment we are carrying out in the present article.

The transition rates in Eq.~(\ref{calPtilde}) could be regarded, in some sense, as a {\it quantum} version of the celebrated Fermi's golden rule.
This should not generate confusion: of course the scattering rates obtained with
the latter are intrinsically quantum in their derivation, but, once computed, they give rise to the Boltzmann equation in~(\ref{BTE}), which describes a \emph{classical} Markov process~\cite{ikeda} for \emph{classical} probabilities; the evolution of the probability density function $f_\lambda$.
Conversely, the transition rates in Eq.~(\ref{calPtilde}) do \emph{not}
describe a classical Markov process, but rather its \emph{quantum} analog: a so called quantum dynamical semigroup~\cite{Lindblad} for the full density matrix. In other words, the dynamics of the entire (i. e., diagonal plus non-diagonal terms) reduced density matrix is determined by a Lindblad-like closed set of equations of the form
\begin{equation}
\label{projLL}
\frac{\mathrm{d} \hat{\rho}^S}{\mathrm{d} t}=
-i  [\hat{\HH}_\circ, \hat{\rho}^S]
- \frac{1}{2}\, \Pp ([\hat{\LL},[\hat{\LL},\hat{\rho}^S]] ) \, ,
\end{equation}
and it is still completely positive and trace invariant.

\section{Application to a prototypical system}
\label{App}
To highlight the critical issues in the CM approximation, let us consider as system $S$ a semiconductor quantum dot (QD) interacting with a solid-state environment $E$ and forced by an external single sound wave source of wavevector ${\bf q}_{\circ}$.
The free Hamiltonian for such a system has the form
\begin{equation}
\label{QDmodel}
\hat{H}_\circ  =  \hat{H}_S \otimes \hat{\mathbbm{1}}_E + \hat{\mathbbm{1}}_S \otimes \hat{H}_E
=\sum_{\alpha}\hbar \omega_\alpha \hat c^\dagger_\alpha \hat c_\alpha +
\sum_{\mathbf{q}}
\hbar\omega_{\mathbf{q}} \,
\hat b^\dagger_{\mathbf{q}} \hat b_{\mathbf{q}} .
\end{equation}
In Eq.~(\ref{QDmodel}), the bosonic operator $\hat b^\dagger_{\bf q}$
($ \hat b^{ }_{\bf q}$) denotes creation (destruction) of a phonon excitation with wavevector ${\bf q}$ and energy $\epsilon_{\bf q}= \hbar\, \omega_{\mathbf{q}}$, while the fermionic operator $\hat c^\dagger_\alpha$ ($\hat c_\alpha$) denotes creation (destruction) of an electron in the state $\alpha$ with energy $\ener_\alpha =\hbar \omega_\alpha$.

The noninteracting (carrier-plus-quasiparticle) basis states $\vert \lambda \rangle$ are given by the tensor product of electronic states $\vert\alpha\rangle$ and quasiparticle states $\vert \{ n_{\bf q} \} \rangle$ corresponding to the occupation numbers $\{n_{{\bf q}}\}$
\begin{equation}
\vert \lambda \rangle = \vert \alpha \rangle \otimes \vert  \{ n_{\bf q}\} \rangle \, ;
\end{equation}
the noninteracting energy spectrum is then the sum of the electronic and quasiparticle energies
\begin{equation}
\hbar\omega_\lambda = \hbar\omega_\alpha  + \sum_{\bf q} \hbar\omega_{\mathbf{q}}\, \hat n_{{\bf q}} \, .
\end{equation}

The perturbation Hamiltonian $ \hat{H}'$ can be written as:
\begin{equation}\label{H_e-qp}
 \hat H' = \hbar \sum_{\bf q}\; \left({\hat \HH}_{\bf q}
 \hat b^{ }_{\bf q} + {\hat \HH}^\dagger_{\bf q}  \hat b^\dagger_{\bf q} \right)
= \hbar({\hat \HH}^{ab} + {\hat \HH}^{em} ) \ ,
\end{equation}
here, ${\hat \HH}_{\bf q} = {\hat \HH}^\dagger_{\bf -q}$ are electronic operators (parametrized by the quasiparticle wavevector ${\bf q}$) acting on the $\alpha$ subsystem only. The two terms in Eq.~(\ref{H_e-qp}) --corresponding to quasiparticle destruction and creation-- describe electronic absorption~(${\hat \HH}^{ab}$) and emission~(${\hat \HH}^{em}$) processes.

In this discussion we write $\hat \HH^i$ as:
\begin{equation}
\begin{aligned}
\hat\HH^i =
\sum_{\alpha_{1}\alpha_{2}}\sum_{\mathbf{q}} \; g_{\alpha_{1}\alpha_{2},{\bf q}}\;
\hat c^\dagger_{\alpha_{1}} \hat c_{\alpha_{2}} \otimes \hat b_{\bf q} +{\rm h.c.}  =\sum_{\alpha_{1}\alpha_{2}}\sum_{\bf q}
\hat c^\dagger_{\alpha_{1}} \hat c_{\alpha_{2}} \otimes \;
\left( g_{\alpha_{1}\alpha_{2},{\bf q}}\,
\hat b_{\bf q}+ g_{\alpha_{2}\alpha_{1},{\bf q}}^*\,
\hat b^\dagger_{\bf q} \right) \, ,
\end{aligned}
\end{equation}
that is, ${\hat \HH}_{\bf q}=
\sum_{\alpha_{1}\alpha_{2}}g_{\alpha_{1}\alpha_{2},{\bf q}}\;
\hat{c}_{\alpha_{1}}^\dagger \hat{c}_{\alpha_{2}}$,
for a generic complex-valued coupling coefficient
$g_{\alpha_{1}\alpha_{2},{\bf q}}$, depending on the wavevector ${\bf q}$.
In particular, within the framework of a deformation potential interaction,
the following form for the
$g_{\alpha_{1}\alpha_{2},{\bf q}}$ applies:
\begin{equation}
\label{gDP}
\begin{aligned}
g_{\alpha_{1}\alpha_{2},{\mathbf{q}}} =
\sqrt{\frac{ \varepsilon^2 \, q^2}{2\,\rho \, V \,
\hbar \omega_{\mathbf{q}}}} \, \int \phi^*_{\alpha_{1}}({\bf r}) \, e^{i\mathbf{q}\cdot {\bf r}} \,
\phi_{\alpha_{2}}({\bf r}) \, \mathrm{d}{\bf r}  = \tilde{g}_{\mathbf{q}}
\braket[ e^{i\mathbf{q}\cdot \mathbf{r}}]{\alpha_{1}}{\alpha_{2}}\,,
\end{aligned}
\end{equation}
where $\varepsilon$ is the deformation potential, $\rho$ is the crystal mass density, and $\phi_{\alpha_{1}}({\bf r})$ [$\phi_{\alpha_{2}}({\bf r})$] is the wavefunction of the QD single electron state $\alpha_{1}$ [$\alpha_{2}$].

The coherent contribution can be taken into account by formally introducing
 a free-part rotation, deriving from $\hat H_\circ$, so that
$\Ll=\Ll^{\rm free}+ \Ll^{\rm scatt}$.
To write{\bf /have} the super-operator in matrix form, we introduce the matrix-to-vector mapping $\rho^S \mapsto \vec{\rhov}^{\,S} $ given by
\begin{equation}
\rhov_i^S= \sum_{\alpha_{1}\alpha_{2}}
\rho^S_{\alpha_{1}\alpha_{2}}\; \delta_{i,(\alpha_{1}-1)n+\alpha_{2}} \, .
\end{equation}
The dynamical equation for the reduced density matrix can then be rewritten as
\begin{eqnarray} \label{rhov20}
\frac{\mathrm{d}}{\mathrm{d} t}  \vec{\rhov}^{\, S} \,(t)=
\Ll \, \vec{\rhov}^{\,S}\,(t) \, .
\end{eqnarray}

In the estimation of $\Ll^{\rm scatt}$, we keep in mind that the projection on the environment degrees of freedom of the operators $\hat{b}^\dagger_{\bf q}  \hat{b}_{\bf q'}$ and
$\hat{b}_{\bf q'}  \hat{b}^\dagger_{\bf q}$ has the following, finite, value
\begin{subequations}\label{B-H}
\begin{align}
\tr[E]{\hat{b}^\dagger_{\bf q}  \hat{b}_{\bf q'} \,\hat{\rho}^{E}}
& = \tilde{N}_{\bf q}\, \delta_{\mathbf{q}\mathbf{q'} }\, , \\
\tr[E]{\hat{b}_{\bf q'}  \hat{b}^\dagger_{\bf q} \,\hat{\rho}^{E}}
& = \left( \tilde{N}_{\bf q} + 1 \right) \delta_{\mathbf{q}\mathbf{q'} } \, ,
\end{align}
\end{subequations}
with $\tilde{N}_{\bf q}$ denoting the average occupation number for state ${\bf q}$.
Moreover, coherent phonon contributions are not accounted for, that is
\begin{equation}
\tr[E]{\hat{b}_{\bf q} \, \hat{\rho}^{E}} = \tr[E]{\hat{b}^\dagger_{\bf q} \, \hat{\rho}^{E}} = 0 \, .
\end{equation}

For the sake of simplicity, let us now assume the effect of the external sound wave source to be dominant in the wavevector distribution of the phonon population. This is indeed a reasonable assumption when, e. g., operating at low lattice temperature. Such a regime may be effectively described by a time-independent diagonal density matrix for the environment, having one element overwhelming the others. It is important to stress that this does not mean a strict one-phonon environment (where any Markov approach would be meaningless), but, on the contrary, a situation in which, while the electron system is forced to very efficiently couple to a certain phonon mode, the latter interacts with the whole lattice.

In this respect, from now on, we limit our attention to the interaction between the system and the externally induced sound wave only. That is, the summations over {\bf q} will be evaluated just for the phonon-mode $\mathbf{q}_\circ$. In particular, this amounts to say that
$\tilde{N}_{\bf q} = \tilde{N} \, q_\circ^3 \, \delta({\bf q}- {\bf q}_{\circ})$.

\subsection{Failure of the conventional Markov approach}
\label{failCM}
In the CM limit, the super-operator $\Ll$ in Eq.~(\ref{matrice}) turns out to be
\begin{equation}
\Ll^{\textrm{scatt}}_{\aaaa}=
 - 2\pi \sum_{\alpha_{3}}
\left(\AAA^{(1)}_{\alpha^\prime_2 \alpha_{3}, \alpha_{3}\alpha_2}\delta_{\alpha_1\alpha^\prime_1}
+\AAA^{(2)}_{\alpha_1 \alpha_{3}, \alpha_{3}\alpha_1^\prime}\delta_{\alpha_2\alpha^\prime_2}
\right) + 2\pi
\left(\AAA^{(1)}_{\alpha^\prime_2 \alpha_2, \alpha_1 \alpha^\prime_1}
+ \AAA^{(2)}_{\alpha^\prime_2 \alpha_2, \alpha_1 \alpha^\prime_1} \right) \, ,
\end{equation}
where
\begin{subequations}
\label{gtilde}
\begin{equation}
{\cal A}^{(1)}_{\aaaa}=
\frac{V}{8 \pi^3} \sum_\pm \int \mathrm{d}^3 \mathbf{q} \,
f_{\aaaa}^{\pm}(\mathbf{q}) \left( \tilde{N}(\mathbf{q}) \pm \frac{1}{2} +\frac{1}{2} \right)
\delta(  \omega(\mathbf{q}) \pm \waan )\, ,
\end{equation}
and
\begin{equation}
{\cal A}^{(2)}_{\aaaa}=
\frac{V}{8 \pi^3} \sum_\pm \int \mathrm{d}^3 \mathbf{q} \,
f_{\aaaa}^{\pm}(\mathbf{q})  \left( \tilde{N}(\mathbf{q}) \pm \frac{1}{2} +\frac{1}{2} \right)
\delta(  \omega(\mathbf{q}) \pm \waap ).
\end{equation}
\end{subequations}
In the latter equations $f^{\pm}$ are defined as
\begin{eqnarray}
\left\lbrace
\begin{array}{ccc}
f_{\aaaa}^{+}(\mathbf{q}) &=&  g_{\alpha_{1}\alpha_{2}}({\bf q})
g_{\alpha_{2}'\alpha_{1}'}^*({\bf q}) \\
f_{\aaaa}^{-}(\mathbf{q}) &=&  g_{\alpha_{2}\alpha_{1}}^*({\bf q})
g_{\alpha_{1}'\alpha_{2}'}({\bf q}) \, ,
\end{array}
\right.
\end{eqnarray}
and $\waan=\omega_{\alpha_{1}}-\omega_{\alpha_{2}}$.

Indeed, using the properties of the Dirac-$\delta$ function
\begin{eqnarray}
\delta(  \omega(\mathbf{q}) \pm \omega_{\alpha_{1} \alpha_{2}} )
\to \frac{\delta(q-q_{\circ})}{|\nabla \omega( \mathbf{q}_{\circ}) |} \, ,
\end{eqnarray}
where $\omega({\bf q}_{\circ}) \equiv \omega_\circ = \mp \omega_{\alpha_{1}\alpha_{2}} $,
we can easily perform the integral in Eq.~(\ref{gtilde}).
Moreover, we may consider our acoustic wave $\mathbf{q}_{\circ}$ in the long-wavelength limit so that we can approximate
\begin{eqnarray}
\omega(\mathbf{q})\approx |\nabla \omega(\mathbf{q})|\, q \, ,
\end{eqnarray}
to have an explicit expression for
the typical integrals ${\cal A}$:
\begin{subequations}
\begin{align}
{\cal A}^{(1)}_{\aaaa}
 = \frac{V |\mathbf{q}_{\circ}|^3 }{8\pi^3\, \omega_{\circ}}
\sum_{\pm} \left[ f^{\pm}_{\aaaa} \left( \tilde{N} \pm \frac{1}{2} +\frac{1}{2} \right)
\delta_{ \omega_{\circ},\,  \pm\, \waan }
\right]
\end{align}
and
\begin{align}
{\cal A}^{(2)}_{\aaaa}
 = \frac{V |\mathbf{q}_{\circ}|^3 }{8\pi^3\, \omega_{\circ}}
\sum_{\pm} \left[ f^{\pm}_{\aaaa} \left( \tilde{N} \pm \frac{1}{2} +\frac{1}{2} \right)
\delta_{ \omega_{\circ}, \, \pm\, \waap } \right]
\end{align}
\end{subequations}
where the functions $f$ (or similarly $g$)
have to be evaluated in $\mathbf{q}_{\circ}$,
and the Kronecker$-\delta$ accounts for energy conservation.

Now, to further focus on our prototypical case of interest, we limit our discussion
to the bound-to-continuum transitions induced by the external source in the electron system of a QD having a single confined state. In this way, labelling the latter as $\vert 1 \rangle$, the only allowed transitions are those into a final state $\vert 2 \rangle$ so that $\omega(\mathbf{q}_{\circ})= \omega_\circ = \omega_{21}$.

The complete super-operator
$\mathbb{L}=\Ll^{\rm free}+\Ll^{\rm scatt}$
may then be written as:
\begin{equation}\label{LL_CM}
\begin{split}
{\Ll} =
\left(
\begin{array}{cccc}
0  & 0 & 0 & 0 \\
 0 & i\omega_{21} & 0 & 0 \\
 0 & 0 & -i\omega_{21} & 0 \\
 0 & 0 & 0 & 0
\end{array}
\right)
 +4\pi \omega_\circ \,\sigma
\left(
\begin{array}{cccc}
-\tilde{N} & 0 & 0 & (\tilde{N}+1) \\
 \zeta \frac{\tilde{N}}{2} &  -\frac{2\tilde{N}+1}{2}&
 \eta \frac{2\tilde{N}+1}{2} &-\zeta\, \frac{\tilde{N}+1}{2}  \\
 \zeta^* \frac{\tilde{N}}{2} &  \eta^* \frac{2\tilde{N}+1}{2} &
 -\frac{2\tilde{N}+1}{2}  &-\zeta^* \frac{\tilde{N}+1}{2}  \\
 \tilde{N} & 0 & 0 & -(N+1)
\end{array}
\right)
\end{split}
\end{equation}
where $\eta=\frac{g_{12}(\mathbf{q}_{\circ})}{g_{21}(\mathbf{q}_{\circ})}$ , $\zeta=\frac{g_{11}(\mathbf{q}_{\circ})-g_{22}(\mathbf{q}_{\circ})}{g_{21}(\mathbf{q}_{\circ})}$, and the dimensionless ratio $\sigma$, defined as
\begin{eqnarray} \label{sigma}
\sigma = \frac{V  }{8\pi^3}\,q_{\circ}^3 \,
\frac{\vert g_{21}(\mathbf{q}_{\circ})\vert^2 }{\omega_\circ^2} ,
\end{eqnarray}
measures the relative strength of the perturbation Hamiltonian.

As can be seen by explicit inspection of the matrix elements of $\Ll$, the equations of motion for the populations $\rho^S_{11}$ and $\rho^S_{22}$ are completely decoupled from those referring to the polarization elements $\rho^S_{12}$ and $\rho^S_{21}=(\rho^S_{12})^*$. On the contrary, the presence of the $\zeta$ and $\eta$ terms lead to a polarization evolution that strongly depends on the populations and can be divergent in some cases.

Defining, for later convenience, the quantities
\begin{subequations} \label{kappas}
\begin{align}
&\kappa = 2\pi (2\tilde{N}+1) \sigma \\
&\kappa' = |\eta |^2 \kappa,
\end{align}
\end{subequations}
we can write the four eigenvalues $\mu_i$ of $\Ll$ as:
\begin{equation} \label{eigen}
\left\{
\begin{array}{rcl}
	\mu_1 &=& 0 \\
	\mu_2 &=& -2\omega_\circ \, \kappa  \\
	\mu_{3,4} &=& \omega_\circ \left(-\kappa \pm \sqrt{\kappa\kappa' -1} \right).
\end{array}
\right.
\end{equation}
The null eigenvalue, $\mu_1$, guarantees the occurrence of a steady-state solution.
It can be shown that the latter corresponds to the vector
\begin{equation} \label{steadyVEC}
\tilde{\rhov}^{\,S} = \frac{1}{2\tilde{N}+1}(\tilde{N}+1,0,0,\tilde{N})^{T} \, ,
\end{equation}
that is, to the $2 \times 2$ matrix
\begin{eqnarray} \label{steady1}
\tilde{{\rho}}^{\,S}= \frac{1}{2\tilde{N}+1}
\left(
\begin{array}{cc}
\tilde{N}+1 & 0 \\
& \\
0 & \tilde{N}
\end{array}
\right) \, .
\end{eqnarray}

We will later show that, in spite of the fact that such a steady state seems so physically sound, the latter may not ever be reached.
As a consequence, the already known pathologies of the CM approach, such as the possible lack of positivity,\cite{D-S} will appear to be much more serious than commonly believed, as they will in fact be proved to show up not only in some initial transient, but also, and heavily, in the long-time/steady state regime.

Coming now to the remaining eigenvalues, by inspection of Eq.~(\ref{eigen}) it is evident that the behaviour of $\mu_3$ (corresponding to the $+$ sign) is crucial. In particular, when
\begin{eqnarray} \label{REmu3}
\sqrt{\kappa\kappa' -1} > \kappa \, ,
\end{eqnarray}
$\Re(\mu_3)$ gets positive and this may lead to a divergence in the polarization terms.
After straightforward manipulation, Eq.~(\ref{REmu3}) may be cast into the following form
\begin{eqnarray} \label{eta}
\kappa^2(|\eta|^2-1)>1 \, .
\end{eqnarray}
The latter is surely not verified if $|\eta|<1$, a case in which the steady state in Eq.~(\ref{steady1}) is always eventually reached. However, when $|\eta|>1$, by rewriting Eq.~(\ref{eta}) with use of Eq.s~(\ref{kappas}), one gets
\begin{eqnarray} \label{crit2}
\sigma >
 \frac{1}{2 \tilde{N}+1 } \cdot
\frac{1}{2\pi \sqrt{|\eta |^2 -1}} \equiv \bar{\sigma} \, .
\end{eqnarray}
that is, $\Re(\mu_3)$ gets positive for $\sigma$ larger than a critical value $\bar{\sigma}$.

An alternative approach to Eq.~(\ref{crit2}) --and more consistent with the
fact that external source intensity is an experimentally controlled degree of freedom--
is to read it in terms of $\tilde{N}$:
\begin{eqnarray} \label{crit}
\tilde{N} > \frac{1}{2} \left(
\frac{1}{\sigma} \,
\frac{1}{2\pi \sqrt{|\eta |^2 -1}} -1 \right) \equiv \bar{N} \, .
\end{eqnarray}
\begin{figure}[hbt]
	\begin{center}
	\includegraphics[scale=0.74]{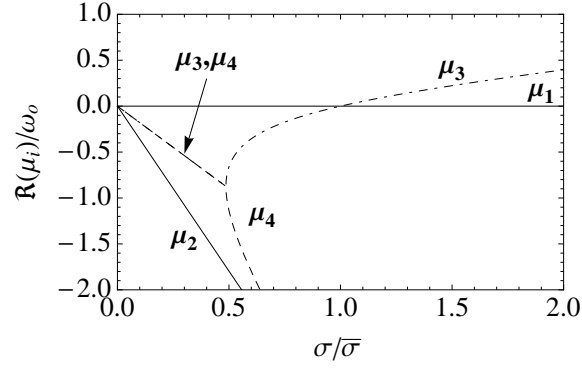}
	\end{center}
	\caption{Real part of the eigenvalues of the super-operator ${\Ll}$,
	in units of $\omega_{\circ}$, as a function of the dimensionless parameter
	$\frac{\sigma}{\bar{\sigma}}$ for the CM approximation.
	Data refer to the case $\vert\eta\vert^2 = 1.31$ and $\tilde{N}=9$, leading to $\bar{\sigma} = 0.01$.}
	\label{SpettL_CM}
\end{figure}

Figure~\ref{SpettL_CM} shows the real part of the four eigenvalues $\mu_i$ of the super-operator $\Ll$ given by Eq.~(\ref{eigen}), in units of $\omega_\circ$, as a function of $\sigma/\bar{\sigma}$, for the case $\eta^2 = 1.31$ (this choice will be motivated later on), that is $|\eta|>1$.  Besides the null eigenvalue, $\mu_1$, and the negative-slope one, $\mu_2$, the detrimental behaviour of $\RR{\mu_3} $, discussed above, is evident. To have a deeper insight into the critical aspects that this may cause, let us now further specify the features of our prototypical system.

To provide a quantitative analysis, the QD three-dimensional confinement potential $V$ is assumed to be made up of a finite square-well profile $V^{\parallel}$, of width $d$, along the growth ($\hat{z}$) direction, and a truncated two-dimensional parabolic potential $V^{\perp}$ in the $xy$ plane. This is a reasonable configuration for state-of-the-art III-V-based self-aggregated QDs.

For the sake of simplicity, and without compromising the analysis we are interested in, in the following we will assume that the factorized form, typical of infinite confinement, remains valid also for our truncated QD. In particular, as single particle bound state $\vert 1 \rangle$, we shall consider the lowest energy one $\langle \mathbf{r}\vert 1 \rangle =
\psi_{01}(\mathbf{r})= \phi_{0}^{\perp}(x,y) \, \phi_{1}^{\parallel}(z)$,
where $\phi_{1}^{\parallel}(z)$ is the ground state of the one-dimensional quantum-well along the $\hat{z}$-axis, centered in $z = 0$ and of width $d$, and $ \phi_{0}^{\perp}(x,y)$ is the ground state of the harmonic oscillator in the $xy$ plane.
Finally, the localization energy (measured from the barrier conduction band minimum) $E_{\rm loc} = \hbar \omega_{\rm loc}$ of state $\psi_{01}(\mathbf{r})$ is set to 7 meV.

To compute the entries of the matrix $\Ll$, namely $\eta$, $\zeta$ and
$\sigma$, we have to calculate the form factors $g_{\alpha\alpha'}({\bf q})$ coupling $\psi_{01}(\mathbf{r})$ to the continuum state $\langle {\bf r}|2\rangle = e^{i \mathbf{k}\cdot \mathbf{r}}/\sqrt{V}$.
In particular, fixing the wave-vectors $\mathbf{q}_{\circ}$ and ${\mathbf{k}}$
along the growth direction ($\hat{z}$), the quantities $\eta$, $\zeta$ and $\sigma $ turn out to be
\begin{subequations}
\begin{align}
\eta =&
\frac{\cos[d(k+q_{\circ})/2]}{\cos[d(k-q_{\circ})/2]} \cdot
\frac{d^2(k-q_{\circ})^2 -\pi^2}{d^2(k+q_{\circ})^2 -\pi^2} \\
\zeta =&
\frac{a d^2 \left(\pi ^2-d^2 (k-q_{\circ})^2\right) q_{\circ}
\sec\left[ d (k-q_{\circ})/2\right] \sin\left[d q_{\circ}/2\right]}
{2 \pi^{3/2}\left(4 \pi^2-d^2 q_{\circ}^2\right)} \\
\sigma =&
\frac{\varepsilon^2 q_{\circ}^5}{a^2 d^2\rho\hbar\omega_\circ^3}
\left\vert
\frac{\cos[d(k-q_{\circ})/2]}{\pi^2 - d^2(k-q_{\circ})^2}
\right\vert^2 \,
\end{align}
\end{subequations}
where $a = (m^* \omega/2 \hbar)^{\frac{1}{2}}$, $\frac{1}{2}m^*\omega^2$ being the paraboloid curvature and $m^*$ the electron effective mass.

Let us now consider an initial condition in which the QD bound level is occupied. This configuration, in the $\{ \vert 1\rangle, \vert 2\rangle \}$ basis, is represented by the following density matrix
\begin{eqnarray} \label{ground}
\rho_g^S=
\left(
\begin{array}{cc}
1 & 0 \\
0 & 0
\end{array}
\right) .
\end{eqnarray}
In the assumed low-temperature limit, and in the absence of any external pumping source, a QD prepared in such a state would indefinitely remain in the latter. This is not the case when the sound wave source is on. At this point, however, some severe pathologies of the conventional approch may show up. Indeed, once the density matrix in Eq.~(\ref{ground}) is rewritten in the basis given by the eigenvectors of $\Ll$, we do expect that the contribution arising from the component $\rho^{S,(\mu_3)}$ may lead to a divergent dynamics when the critical conditions of Eq.~(\ref{REmu3}) are met.

\begin{figure}
	\begin{center}
	\includegraphics[scale=.74]{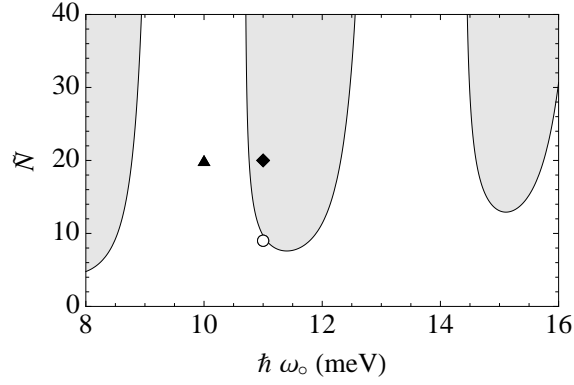}
	\end{center}
	\caption{Partition of the ($\tilde{N}$, $\hbar\omega_\circ$)
	parameter space produced by the $\tilde{N} = \bar{N}$ equation.
	The triangle (diamond) highlights the configuration in which the phonon energy
    $\hbar \omega_\circ$ amounts to 10 meV (11 meV), with $\tilde{N}=20$.
	The circle corresponds to the case $\hbar\omega_\circ=11$ meV
	and $\tilde{N}=9$ ($\bar{\sigma} = 0.01$). }
	\label{EsisteSempre}
\end{figure}
A convenient way to picture out such a non trivial behaviour is given by Fig.~\ref{EsisteSempre}. In the latter, $\bar{N}$ is plotted as a function of the phonon energy $\hbar\omega_{\circ}$. This produces a partition of the ($\tilde{N}$, $\hbar\omega_{\circ}$) parameter plane into distinct regions, according to whether the condition in Eq.~(\ref{crit}) is satisfied or not. In particular, there appear a ``nonphysical region'' (marked in grey), where the density matrix eigenvalues diverge, and a region in which, on the contrary, Eq.~(\ref{crit}) is not satisfied therefore guaranteeing the occurrence of the expected steady-state solution.

To better focus on such a density matrix dynamics, let us consider two diverse setups, differing in the energy, $\hbar\omega_{\circ}$, of the incident phonon beam: a first one in which $\hbar\omega_\circ = 11$ meV, and a second one with $\hbar\omega_{\circ}$= 10 meV. For the former, marked by a diamond in Fig.~\ref{EsisteSempre}, the critical region shows up for $\bar{N} \gtrsim 10$; this is not the case of the latter, marked by a triangle, which falls in the `safe' region. Indeed, despite of the fact that the two phonon energies are quite similar, the effects on the system evolution are dramatically different, as we are showing in the following.

In particular, considering the time evolution of the density matrix elements $\rho^S_{ij}$ when our electronic system is initially prepared in the $\rho_g^S$ state and $\hbar\omega_{\circ}$= 10 meV, the population terms, $\rho^S_{11}$ and $\rho^S_{22}$, are found to converge to a steady-state value in a sub-picosecond timescale, while the polarization contributions, $\rho^S_{12}$ and $\rho^S_{21}$, decay exponentially to zero, as expected.

\begin{figure}[hbt]
	\begin{center}
	\includegraphics[scale=.77]{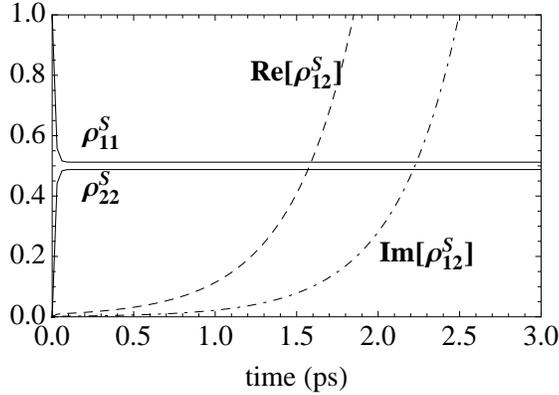}
	\end{center}
	\caption{Time evolution of the system density matrix elements, $\rho^S_{ij}$, in the
    $\{\vert 1 \rangle, \vert 2 \rangle\}$ basis, when the QD is initially prepared in
     the $\vert 1 \rangle$ state. Continuous lines refer to
     the populations $\rho^S_{11}$ and $\rho^S_{22}$;
     the dashed (dot-dashed) line corresponds to the real (imaginary) part
     of the polarization term $\rho^S_{12}$.
    The phonon frequency is $\hbar\omega_{\circ}=11$ meV, $\sigma=0.0092$,
    and $\tilde{N}=20$.
    }
	\label{11meV}
\end{figure}

Let us now move to the case in which $\hbar \omega_\circ = 11$ meV. As we may expect from the evidences given in the previous Section, some critical aspects should appear in the dynamics of $\rho^S$ making the latter no longer positive definite. Indeed, as shown in Fig.~\ref{11meV}, though the population time evolution (continuous lines) still appears somehow reasonable (i.e., the trace is preserved), the polarization one (dashed and dot-dashed lines) is unphysically divergent.
As highlighted before, the origin of these pathologies is in the lack of temporal symmetry in the construction of the CM approximation; namely, in the double commutator expression in Eq.~(\ref{LvN-eff}).

\subsection{Success of the alternative Markov procedure}
\label{AppNewM}
We are now interested in comparing the results of Subsection~\ref{failCM} with those derived from above described alternative approach to the Markov approximation; we will show that the previously mentioned pathologies of CM are now brilliantly overcome.

First of all, performing the trace over the environment degrees of freedom, Eq.~(\ref{projLL})
turns into
\begin{equation}
\label{projLL2}
\frac{\mathrm{d}}{ \mathrm{d} t} {\hat{\rho}}^S
= \sum_{\mathbf{q}\, ,\pm}
\Big( \tilde{N}_{\bf q}+\frac{1 }{ 2} \pm \frac{1 }{ 2}\Big)
\left( - \frac{1}{2} \left\{ {\cal L}^{\pm \dagger}_{\bf q}
{\cal L}^{\pm}_{\bf q},  \hat{\rho}^{S} \right\} + {\cal L}^{\pm}_{\bf q}
 \hat{\rho}^{S}
{\cal L}^{\pm \dagger}_{\bf q}
\right)
\end{equation}
with
\begin{equation}\label{calLpm}
{\cal L}^\pm_{\bf q} =
\left(\frac{1 }{ 2\pi\overline{t}^2}\right)^\frac{1 }{ 4}
\int_{-\infty}^\infty \mathrm{d}t \;{H}^\pm_{\bf q}(t)\; e^{-\frac{t^2 }{ 4\overline{t}^2}} \ ,
\end{equation}
where
\begin{equation}
{H}^\pm_{\bf q}(t) = \sum_{\alpha_{1} \alpha_{2}}
\hat{c}^\dagger_{\alpha_{1}} \hat{c}_{\alpha_{2}} \,
g^{\pm}_{\alpha_{1} \alpha_{2} , {\bf q}} \,
e^{-i \,(\omega_{\alpha_{1} \alpha_{2}} \pm \omega_\circ)\, t} \, ,
\end{equation}
$g^{-}_{\alpha_{1} \alpha_{2} , {\bf q}}=g_{\alpha_{1} \alpha_{2} , {\bf q}}$,
$g^{+}_{\alpha_{1} \alpha_{2} , {\bf q}}=g^{*}_{\alpha_{2} \alpha_{1}, {\bf q}}$,
and $\hat{\rho}^{E}$ is again the environment density matrix.

With the above explicit expression of the interaction Hamiltonian, it is easy to perform the time integrals in Eq.~(\ref{calLpm}), since the latter only involve products of gaussian functions and plane waves. Equation~(\ref{projLL2}) is then cast into the form of Eq.~(\ref{matrice}), however, contrary to the CM result, the novel $\PP$ terms are now given by\cite{propGauss}
\begin{equation}
\label{nuova1}
\PP_{\aaaa} = 2\pi\,
e^{-\frac{\left(\waa{1}-\waa{2}\right)^2}{8 \bar{\omega}^2}}
\sum_{\mathbf{q}\, ,\pm}
\left( \tilde{N}_\mathbf{q} +\frac{1}{2}\pm \frac{1}{2} \right)
g_{\alpha_1\alpha_1',\mathbf{q}}\,g^{*}_{\alpha_2\alpha_2',\mathbf{q}}
\,\frac{1}{\sqrt{2\pi} \bar{\omega}}
\, e^{ -\frac{\left(\!\frac{\waa{1}+\waa{2}}{2}
\pm\omega_{\mathbf{q}}\!\!\right)^2}{2 \bar{\omega}^2 }} \, .
\end{equation}

In the completed-collision limit ($\bar{\omega}\to 0$),
that is, when the broadening due to the the gaussian correlation function is null,
the first exponential term in Eq.~(\ref{nuova1}) results into
a Kronecker $\delta$-function, while the second one produces a Dirac $\delta$-function as follows
\begin{equation}
\label{genNM}
\PP_{\aaaa}  =   2\pi\, \delta_{\waa{1},\,\waa{2}}
\frac{V}{8 \pi^3}\! \sum_{\pm}\! \int\!\! \mathrm{d} \mathbf{q}\!
\left(\!\! \tilde{N}_\mathbf{q} +\frac{1}{2}\pm \frac{1}{2}\! \right)
g_{\alpha_1\alpha_1'}(\mathbf{q})\,g^{*}_{\alpha_2\alpha_2'}(\mathbf{q})
\,\delta\!\left(\frac{\waa{1}+\waa{2}}{2} \pm\omega_{\mathbf{q}}\right).
\end{equation}
While the former forces the system to have a discrete spectrum --the only one considered
in the exiting literature, up to now-- the latter corresponds to energy conservation.
It is straightforward to verify that the usual semiclassical rates $\PP_{\alpha\alpha^\prime}$ given by the Fermi's golden rule may be obtained from Eq.~(\ref{genNM}) when $\alpha_1=\alpha_2=\lambda$ and $\alpha_1^\prime=\alpha_2^\prime=\lambda^\prime$.

For a more quantitative analysis, let us now focus on the prototypical system considered in Section~\ref{App}, that is a semiconductor QD interacting with a collimated and monoenergetic source of phonons of wavevector ${\bf q}_\circ$. The latter, in particular, only couples the QD bound state $\vert 1 \rangle$ to a specific continuum one, $\vert 2 \rangle$.
Substituting for the explicit expression
$\tilde{N}_{\bf q}= \tilde{N}\, q_\circ^3 \, \delta({\bf q}- {\bf q}_{\circ})$,
the generalized scattering rates become
\begin{equation}
\PP_{\aaaa} = 2\pi\, \delta_{\waa{1},\,\waa{2}} \,
 \frac{V q_{\circ}^3}{8 \pi^3 \omega_{\circ} }
\sum_{\pm}  \left( \tilde{N} +\frac{1}{2}\pm \frac{1}{2} \right)
g_{\alpha_1\alpha_1'}\,g^*_{\alpha_2\alpha_2'}
\; 2\,\delta_{ \waa{1}+\waa{2}\,, \mp\, 2 \omega_{\circ}} \, ,
\end{equation}
where $\alpha = 1,2$.
In our approach, the markovian super-operator ${\Ll} = {\Ll}^{\rm free} + {\Ll}^{\rm scatt}$ may now be written in matrix form. In particular, the free part, ${\Ll}^{\rm free}$, is identical to the one obtained in Subsection~\ref{failCM},
\begin{equation}\label{LfreeNM}
{\Ll}^{\rm free}\! \!  =
\left(
\begin{array}{cccc}
0 & 0 & 0 & 0 \\
0 & i\omega_{21} & 0 & 0 \\
0 & 0 & -i\omega_{21} & 0 \\
0 & 0 & 0 & 0
\end{array}
\right) \, ,
\end{equation}
while ${\Ll}^{\rm scatt}$ now becomes
\begin{equation}\label{LscattNM}
{\Ll}^{\rm scatt}\! \!  = 4 \pi \omega_{\circ}\,
\sigma \left(
\begin{array}{cccc}
-\tilde{N} & 0 & 0 & \tilde{N}+1 \\
0 & -\frac{2\tilde{N}+1}{2} & 0 & 0 \\
0 & 0 & -\frac{2\tilde{N}+1}{2} & 0 \\
\tilde{N}  & 0 & 0 & -\tilde{N}-1
\end{array}
\right) \, .
\end{equation}
with $\sigma $ as given in Eq.~(\ref{sigma}).

Let us now compare Eq.~(\ref{LscattNM}) with the scattering part of Eq.~(\ref{LL_CM}) in the CM-limit case.
First of all, when $|g_{21}|=0$, the scattering super-operator vanishes in both of them. However, in Eq.~(\ref{LscattNM}) this result simply derives from the fact that ${\Ll}^{\rm scatt}$ depends {\it only} on $g_{21}$, through $\sigma$. The lack of any coupling between the populations and polarizations dynamics in Eq.~(\ref{LscattNM}) leads to a time evolution of $\rho$ that, contrarily to the CM approximation, does not have any $T_3$ contribution.\cite{PRB}
Moreover, although the first and last rows of the ${\Ll}^{\rm scatt}$ matrix are identical in the two approaches, the dynamics of the polarizations is totally different. This feature is clearly evident when one diagonalizes $\Ll $.
Indeed, the eigenvalues of the latter are
\begin{equation}\label{eigs}
\left\{
\begin{array}{rcl}
\mu_1 &=& 0\\
\mu_2 &=& -\omega_{\circ} (\kappa +i)\\
\mu_3 &=& (\mu_2)^* \\
\mu_4 &=& -2\omega_{\circ} \kappa
\end{array}
\right.
\end{equation}
where $\kappa$ is related to $\sigma$ by
$\kappa= 2 \pi (2\tilde{N}+1) \sigma $.

\begin{figure}[htb]
\includegraphics[scale=.74]{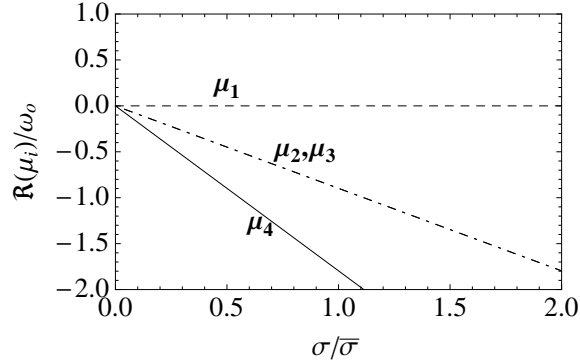}
	\caption{Real parts of the eigenvalues $\mu_i$ of the
	super-operator $\Ll$, plotted as a function of $\sigma/\bar{\sigma}$
	to allow for a direct
    comparison with Fig.~\ref{SpettL_CM}. }
\label{AutoL}
\end{figure}

Figure~\ref{AutoL} presents the eigenvalues $\mu_i$ just obtained as a function of $\sigma/\bar{\sigma}$. This choice is made to directly compare these results with the ones showed in Fig.~\ref{SpettL_CM}; in particular, $\bar{\sigma}$ has been defined in Eq.~(\ref{crit2}), within the CM-limit framework, and is here set to the value $\bar{\sigma}=0.01$.
We stress that the real part of the eigenvalues in Eq.~(\ref{eigs}) is always non-positive, regardless of the value of the $g_{ij}$'s; the zero eigenvalue corresponds to the steady state solution
\begin{eqnarray} \label{sstate}
\tilde{\rho}^{\,S}= \frac{1}{2\tilde{N}+1}
\left(
\begin{array}{cc}
\tilde{N}+1 & 0 \\
& \\
0 & \tilde{N}
\end{array}
\right).
\end{eqnarray}
identical to the result in Eq.~(\ref{steady1}), obtained with the CM approach.

Let us now turn to Eq.~(\ref{rhov20}).
By inspection of Eq.s~(\ref{LfreeNM}) and~(\ref{LscattNM}) $\Ll$ contains an inner $2 \times 2$ diagonal block, relative to the polarizations dynamics, while the evolution of the population is dictated by coefficients identical to those of the CM approach. This strongly simplifies the problem, allowing for an analytical solution. The vector $\vec{\rhov}^{\,S}$  is then
\begin{equation}
\label{RhoTempo}
\vec{\rhov}^{\,S}(t)=
\left(
\begin{array}{rl}
\rho_{11}^{S,(\mu_1)}+
\left( \rho^S_{11}(0)-\rho_{11}^{S,(\mu_1)}\right) & \!\! e^{-2\kappa\,\omega_{\circ} t} \\
\rho^S_{12}(0) &\!\! e^{-(\kappa + i)\omega_{\circ} t}  \\
\rho^S_{21}(0) & \!\! e^{-(\kappa - i)\omega_{\circ} t} \\
\rho_{22}^{S,(\mu_1)} -
\left(\rho^S_{11}(0)-\rho_{11}^{S,(\mu_1)}\right) & \!\! e^{-2\kappa\,\omega_{\circ} t}  \\
\end{array}
\right).
\end{equation}
where $\rho^S_{ij}(0)$ are the elements of the $2 \times 2$ density matrix $\rho^S$ at time $t = 0$. It is important to notice that $\vec{\rhov}^{\,S}$ can now be read as a sum of the steady-state, i.e., \mbox{$t$-independent}, solution $\tilde{\rho}^{\,S}$ given in Eq.~(\ref{sstate}), and a contribution $\delta\rho(t)$ which decays exponentially and contains the characteristic
times $T_1=(2\kappa\,\omega_{\circ})^{-1}$ and
$T_2=2\,T_1$,
\begin{eqnarray}
\rho^{S}(t)=\tilde{\rho}^{\,S}+\delta\rho(t) \,.
\end{eqnarray}

In other words, whatever being the initial condition $\rho^S(0)$,
the density matrix reaches the equilibrium when the real parts of the
exponential functions in Eq.~(\ref{RhoTempo}) approach zero; the intensity of
the $\delta\rho(t)$ term is proportional
to the difference between the initial condition and
the steady state solution.
Moreover, contrarily to the CM case, where severe instabilities may result
from a \mbox{\emph{diagonal}} perturbation produced in the system steady-state,
here the form of $\delta\rho(t)$ is so
that the latter does not affect the polarization dynamics at any time.

\begin{figure}[htb]
	\begin{center}
	\includegraphics[scale=0.72]{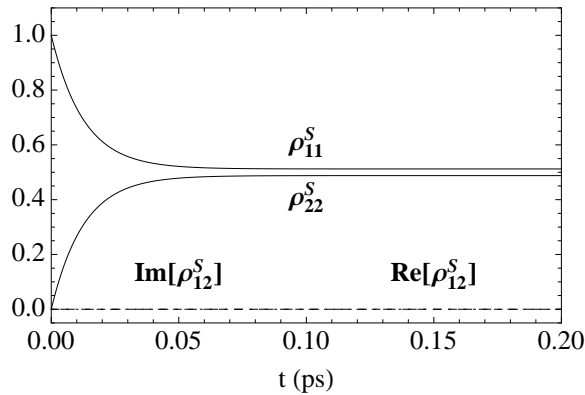}
	\end{center}
	\caption{Time evolution of the population and polarization terms of $\rho^S$, in the
    $\{\vert 1 \rangle, \vert 2 \rangle\}$ basis, when the QD is initially prepared in
    the $\vert 1 \rangle$ state. Continuous lines refer to
    the populations $\rho^S_{11}$ and $\rho^S_{22}$;
    the dashed (dot-dashed) line corresponds to the real (imaginary) part
    of the polarization term $\rho^S_{12}$.
    The phonon frequency is $\hbar\omega_{\circ}=11$ meV, $\sigma=0.0092$,
    and $\tilde{N}=20$.}
	\label{rhoG11meV}
\end{figure}

To be more quantitative, and also to compare the results of Subsection~\ref{failCM} with the dynamics given by our alternative approach, let us consider a density matrix $\rho^S$ initially prepared as in Eq.~(\ref{ground}) and the configuration marked by a diamond in Fig.~\ref{EsisteSempre}.
The latter was a critical situation for the CM method that produced the divergent behaviour in the polarization dynamics shown in Fig.~\ref{11meV}. The situation resulting from our approach is completely different, being stable and physically meaningful. In particular, the time evolution of $\rho^S$ given by Eq.s~(\ref{LfreeNM}) and (\ref{LscattNM}) produces the results shown in
Fig.~\ref{rhoG11meV}. The population terms $\rho^S_{11}$ and $\rho^S_{22}$ decay exponentially from the initial values 1 and 0, respectively, to the steady state ones $\tilde{\rho}^S_{11}\simeq 0.51$ and $\tilde{\rho}^S_{22}\simeq 0.49$, which are identical to those found in the case reported in Fig.~\ref{11meV} for the CM approach. The main discrepancy between the two methods appears when considering the behaviour of the polarization terms: following our prescriptions, and conversely to the divergent CM result, the latter {\it remain} null, as they initially are.

\section{Summary and Conclusions}\label{s-SC}
We have presented and discussed a general density-matrix description of energy-dissipation and decoherence phenomena in open quantum systems, able to overcome the intrinsic limitations of the conventional Markov approximation.
More specifically, contrary to the usual single-particle correlation expansion, we have investigated the effect of the Markov limit, in a fully operatorial approach, before applying any
reduction procedure. This has allowed us to better identify the general properties of the scattering
super-operators entering our effective quantum-transport theory and to propose an alternative adiabatic scheme that does not threaten positivity at any time. The key idea of our approach consists in performing the temporal symmetrization and coarse graining of the scattering term in the Liouville-von Neumann equation, prior to the reduction over the environment degrees of freedom. The result is a robust treatment of energy-dissipation and dephasing phenomena in state-of-the-art semiconductor quantum devices, able to describe a genuine Lindblad-like evolution that also recovers the Fermi's golden rule features in the semiclassical limit.

Applications to the prototypical case of a semiconductor quantum dot exposed to a single-phonon source are discussed, highlighting the success of our formalism with respect to the critical issues of the conventional Markov limit. In particular, we have shown very important intrinsic limitations
of the CM approach when employed for large time and/or steady state analysis, producing severe instabilities that show up in divergent polarization terms.

In the present article we focused on the dissipative part of the scattering super-operator,
deliberately neglecting the energy level renormalization. Indeed, as can be checked both analytically and numerically, the effect of the latter is to worsen the pathologies of the conventional Markov limit, rather than to remedy for them. Moreover, the presence of energy renormalization effects in our proposed scheme would result into a coherent term in the dynamical equation for the global density matrix, which does not affect positivity in any case.

We stress that our formulation significantly generalizes preexisting theories, as it gives a positive dynamics for a considerably large class of projections (i.e., ways to chose the subsystem), irrespective of the subsystem's dimension or spectral properties. This allows, on one side, to
investigate subsystems with both discrete and continuous spectra --a feature largely shared by mesoscale electronic and opto-electronic quantum devices-- and suggests, on the other side, a
new way to treat electrical contacts for quantum devices, thus revealing an interesting potential for an alternative quantum transport formalism.

Finally, it is imperative to stress that, in the presence of a strong system-environment interaction, the adiabatic decoupling investigated so far needs to be replaced by more realistic treatments, expressed via non-Markovian integro-differential equations of motion~\cite{RMP} (i.e., with ``memory effects''). Again, while for purely atomic and/or photonic systems it is possible to identify effective non-Markovian evolution operators,\cite{nonmarkovian} for solid-state quantum devices
this is still an open problem.

%\section*{Acknowledgments}

\newpage


\begin{thebibliography}{xx}
\bibitem{renorm}
Indeed, one could always reconsider the condition $\Pp\left([\hat H'(t),\Pp {\hat \rho}]\right)= 0$  by assigning the first order quantity $\Pp(\hat H')$ to the deterministic part of the Hamiltonian $\hat H_\circ$, which however could then be more difficult to solve.
\bibitem{propGauss}
The exponent of the gaussian functions has been rewritten as
$\left\{ \frac{-(a\pm \omega)^2-(b\pm\omega)^2}
{4 \bar{\omega}^2 } \right\}
= \left\{ -\frac{\left(\frac{a+b}{2} \pm\omega \right)^2}
{2 \bar{\omega}^2 } - \frac{(a-b)^2}{8 \bar{\omega}^2 } \right\}$.
\end{thebibliography}
\end{document}